\documentclass[prl,twocolumn,superscriptaddress,showpacs]{revtex4-1}

\usepackage{graphicx}

\begin{document}

\title{Superconductivity tuned by the iron vacancy order in K$_{\bf x}$Fe$_{\bf 2-y}$Se$_2$}


\author{Wei Bao}
\email{wbao@ruc.edu.cn}
\affiliation{Department of Physics, Renmin University of China, Beijing 100872, China}
\author{G. N. Li}
\affiliation{NIST Center for Neutron Research, National Institute of Standards
and Technology, Gaithersburg, MD 20899, USA}
\affiliation{Institute of Physics, Chinese Academy of Sciences, Beijing 100190, China}
\author{Q. Huang}
\affiliation{NIST Center for Neutron Research, National Institute of Standards
and Technology, Gaithersburg, MD 20899, USA}
\author{G. F. Chen}
\author{J. B. He}
\author{D. M. Wang}
\affiliation{Department of Physics, Renmin University of China, Beijing 100872, China}
\author{M. A. Green}
\author{Y. Qiu}
\affiliation{NIST Center for Neutron Research, National Institute of Standards
and Technology, Gaithersburg, MD 20899, USA}
\affiliation{Department of Materials Science and Engineering,
University of Maryland, College Park, MD 20742, USA}
\author{J. L. Luo}
\affiliation{Institute of Physics, Chinese Academy of Sciences, Beijing 100190, China}
\author{M. M. Wu}
\affiliation{Neutron Scattering Laboratory, China Institute of Atomic Energy, Beijing 102413, China}

\date{\today}

\begin{abstract}
Combining in-depth neutron diffraction and systematic bulk studies, we discover that the $\sqrt{5}\times\sqrt{5}$ Fe vacancy order with its associated block antiferromagnetic order is the ground state, with varying occupancy ratio of the iron $16i$ and vacancy $4d$ sites, across the phase-diagram of K$_{x}$Fe$_{2-y}$Se$_2$. The orthorhombic order with one of the four Fe sites vacant appears only at intermediate temperature as a competing phase. The material experiences an insulator to metal crossover when the $\sqrt{5}\times\sqrt{5}$ order has highly developed. Superconductivity occurs in such a metallic phase.
\end{abstract}


\maketitle 

The recently discovered alkali metal ($A$=K, Rb, Cs, Tl-K or Tl-Rb) intercalated iron selenide superconductors \cite{C122924,C123637,C125236,C125525} achieve much higher a transition temperature ($T_c\sim 30$ K) than in previous iron selenide (11) superconductors ($T_c\le 14$ K) \cite{A072369} at ambient pressure. They were initially thought to be isostructural to BaFe$_2$As$_2$, and the Fe valence at 1.6+ implied by chemical formulas $A_{0.8}$Fe$_2$Se$_2$ seems supported by ARPES studies \cite{C125980}. However, vacancies have been intentionally introduced in the Fe plane in the Tl-containing superconductors \cite{C125236}, and various Fe vacancy orders have also been directly observed in the K intercalated materials in transmission electron microscopy (TEM) study \cite{D012059}. Moreover, all of the new superconductors have been shown via neutron and x-ray diffraction refinement method to be close to the chemical composition $A_{0.8}$Fe$_{1.6}$Se$_2$ or $A_2$Fe$_4$Se$_5$ (245), to experience the same $\sqrt{5}\times\sqrt{5}$ type of Fe vacancy order-disorder transition, and to order in the same large-moment block antiferromagnetic structure \cite{D020830,D014882,D022882}. Therefore, the Fe vacancies are introduced in the 245 superconductors to ensure the Fe valence to be close to 2+ \cite{D014882}, as in the 11 iron selenide superconductors \cite{A092058}. 

Recently a series of K$_{x}$Fe$_{2-y}$Se$_2$ samples have been synthesized and their transport property ranging from superconducting to insulating systematically studied \cite{D010789}.
Two of the superconducting samples in the series are the subject of previous refinement studies \cite{D020830,D014882}. Here we focus on an insulating sample K$_{0.99}$Fe$_{1.48}$Se$_2$ to contrast its physical properties with the superconducting samples. The key difference is in the {\em degree} of the $\sqrt{5}\times\sqrt{5}$ Fe vacancy order at low temperature. The imperfect $\sqrt{5}\times\sqrt{5}$ order in K$_{0.99}$Fe$_{1.48}$Se$_2$ introduces severe impurity scattering and renders the material insulating. Similarly, Anderson weak localization, albeit by spin-glass low energy scattering, has been shown previously to destroy bulk superconductivity in the 11 superconductors \cite{C035647}. Above $T^*$ at a structural transition, the $\sqrt{5}\times\sqrt{5}$ vacancy order is partially replaced by an orthorhombic vacancy order of a $\sqrt{2}\times\sqrt{2}$ or $2\sqrt{2}\times\sqrt{2}$ supercell. This transition was previously misinterpreted as a N\'{e}el transition \cite{C125236}, which led to an erroneous quantum critical theory. The insulating samples also maintain the large-moment block antiferromagnetic order to high temperature. A phase diagram for the K$_{x}$Fe$_{2-y}$Se$_2$ system is provided.

The single crystal samples were grown by the Bridgman method \cite{D010789}, and the composition was determined by inductively coupled plasma (ICP) analysis and neutron powder diffraction refinement. Resistivity measurements were carried out with a physical property measurement system (PPMS). Magnetic susceptibility was measured with the PPMS below 300 K at RUC, and above 300 K using the oven option at IOP.
10 g of K$_{0.99}$Fe$_{1.48}$Se$_2$ crystals were ground to powder in a He atmosphere and then sealed with He exchange gas in a vanadium can for study using the BT-1 high-resolution neutron powder diffractometer at NCNR. The Cu (311) and Ge(311) monochromators were used to obtain neutron beams at wavelength 1.5403 and 2.0783 $\AA$, respectively. Horizontal collimators 60$^{\prime}$-20$^{\prime}$-7$^{\prime}$ were used. Neutron data were collected with steps of 0.05$^o$ in the 2$\theta$ range 3$^o$-168$^o$. Sample temperature was controlled by a high temperature Displex closed cycle refrigerator.
The nuclear and magnetic structures were refined using the GSAS program. After picking out non-reacting Fe plates during grinding, 2.50(6)\% Fe impurity was found in the powder sample and was accounted for in the refinement. 

\begin{table}[b!]
\caption{Refined structure parameters at 580 K for K$_{0.99(1)}$Fe$_{1.48(1)}$Se$_2$ in the ThCr$_2$Si$_2$ structure.
Space group $I4/mmm$ with $a=3.9999(1)\AA$, $c=13.918(5)\AA$ and $V=222.70(2)\AA^3$.
$R_p=4.49$\%, $R_{wp}=5.54$\% and $\chi^2=1.582$.}
\label{tab1}
\begin{tabular}{cccccccc}
\hline \hline
 atom  & site & $x$  & $y$  & $z$ & $B_{11}$=$B_{22}$ ($\AA^2$) & $B_{33}$ ($\AA^2$) & $n$\\
    \hline
  K & 2$a$ & 0 & 0             & 0             & 3.6(2)  & 3.4(3) & 0.99(1) \\
 Fe & 4$d$ & 0 & $\frac{1}{2}$ & $\frac{1}{4}$ & 2.43(6) & 2.3(1) & 0.740(5) \\
 Se & 4$e$ & 0 & 0             & 0.3546(1)     & 2.52(5) & 2.8(1) & 1  \\  
\hline \hline
    \end{tabular}
\end{table}
 
\begin{figure}
\includegraphics[width=.6\columnwidth]{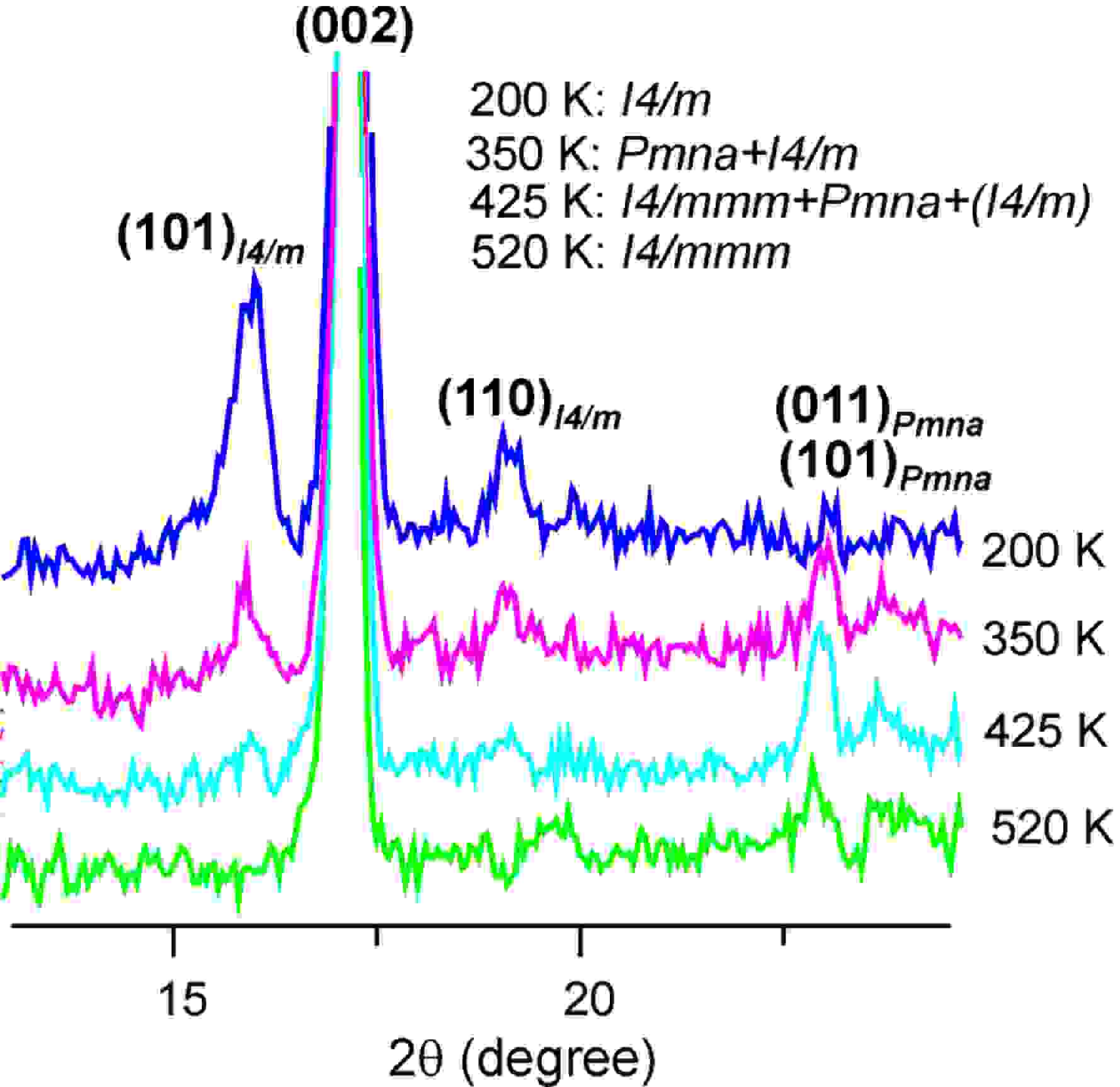}
\includegraphics[width=.8\columnwidth]{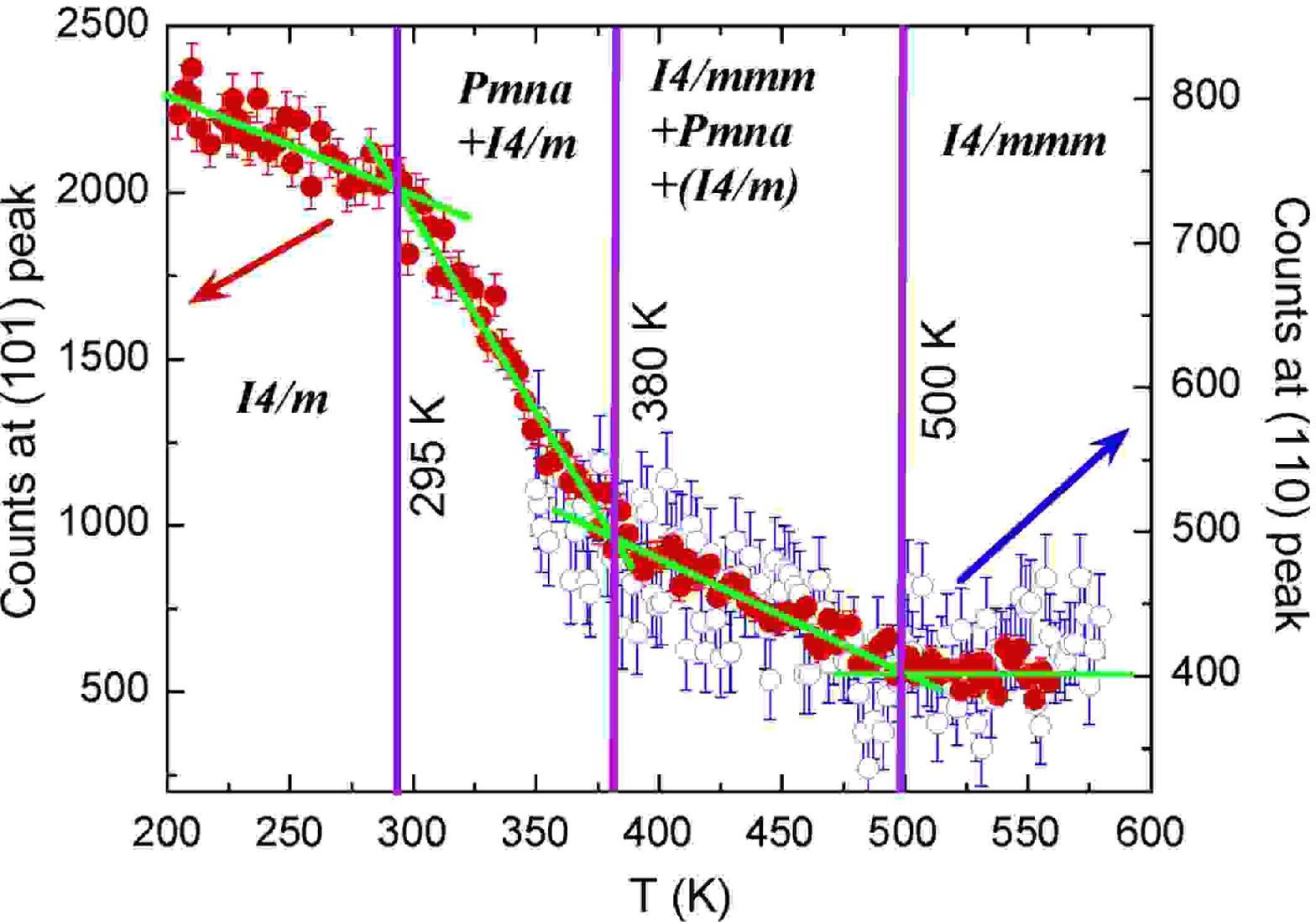}
\vskip -.3cm
\caption{(a) A segment of the neutron powder diffraction spectrum of K$_{0.99}$Fe$_{1.48}$Se$_2$, showing the appearance of new Bragg peaks (101)/(011), (110) and (101) with decreasing temperature as the $Pmna$ and $I4/m$ Fe vacancy orders and the block antiferromagnetic order break the $I4/mmm$ symmetry, respectively. (b) The structural (110) (open blue circles) and magnetic (101) (solid red circles) Bragg peaks as the order parameter of the $I4/m$ phase, measured on warming. Below $T_{S,N}\approx 500$ K and above $T^*\approx 295$ K, the $Pmna$ and $I4/m$ Fe vacancy orders coexist. The disordered $I4/mmm$ phase extends down to $\sim 380$ K. }
\label{fig1}
\end{figure}

Fig.~1(a) highlights neutron powder diffraction pattern in a limited $2\theta$ range at the selected temperatures. At 520 K, there exists the (002) Bragg peak of the tetragonal ThCr$_2$Si$_2$ structure. The refined structure parameters from the whole pattern at 580 K are listed in
Table~\ref{tab1}, showing that the K site is almost fully occupied and that 74\% Fe and 26\% vacancy are randomly mixed at the 4$d$ Fe site. Note that the difference in the ThCr$_2$Si$_2$ structure of KFe$_{1.48}$Se$_2$ and KFe$_{2}$Se$_2$ shown in Fig.~2(a) cannot be detected without a refinement of the Fe occupancy $n$. Many mistakes in misidentifying the sample composition in current literature is due to this fact. 

At 200 K, the diffraction pattern in Fig.~1(a),
similar to that for the superconducting K$_{0.83(2)}$Fe$_{1.64(1)}$Se$_2$ sample \cite{D020830}, shows the (110) Bragg peak due to the $\sqrt{5}\times\sqrt{5}$ Fe vacancy order and the (101) due to the block antiferromagnetic order. However, the 26\% Fe vacancy in the current sample exceeds the 20\% of the perfect $\sqrt{5}\times\sqrt{5}$ order shown in Fig.~2(b).
This leads to less than full occupancy of the 16$i$ Fe2 site and substantial occupancy of the vacant 4$d$ Fe1 site of the tetragonal $I4/m$ structure, as the refinement results at 50 and 295 K in Table~\ref{tab2} show. Thus, unlike the almost perfect vacancy order in the superconducting samples which folds the Brillouin zone and contributes little to electric resistance, there exists substantial Fe site disorder in the low temperature phase of K$_{0.99}$Fe$_{1.48}$Se$_2$ which renders the ground state insulating with a transport gap of $\sim 85$ meV [Fig.~\ref{fig3}(b)].

\begin{table*}[tb!]
\caption{Refined structure parameters for K$_{0.99}$Fe$_{1.48}$Se$_2$ in the $\sqrt{5}\times\sqrt{5}$ structure with space group $I4/m$. The $B$ factor was constrained to be the same for the same elements in the final refinement.}
\label{tab2}
\begin{center}
\begin{tabular}{cc|cccccc|cccccc} \hline\hline
\multicolumn{2}{c}{} &
\multicolumn{6}{|c}{50 K } &
\multicolumn{6}{|c}{ 295 K }  \\ 
\multicolumn{2}{c}{} &
\multicolumn{6}{|c}{a=8.7941(3)\AA, c=13.8062(6) \AA, V=1067.72(9) \AA$^{3}$} & 
\multicolumn{6}{|c}{a=8.8389(7)\AA, c=13.8909(1) \AA, V=1085.26(8) \AA$^{3}$}  \\ 
\multicolumn{2}{c}{} &
\multicolumn{6}{|c}{$R_p=7.89\%,~wR_p=9.63\% $, $\chi^2=1.429$} &
\multicolumn{6}{|c}{$R_p=4.74\%,~wR_p=5.68\% $, $\chi^2=2.021$} \\ \hline
Atom&site &x &y &z &B(\AA$^{2}$) & n & M$_z(\mu_B)$ &
x &y &z &B(\AA$^{2}$) &  n  & M$_z(\mu_B)$\\ \hline 
K1 & 2a & 0 &0 & 0 & 0.72(1) & 1 & &0 &0 & 0 & 2.4(1) & 1 & \\
K2 & 8h & 0.395(2) &0.197(3) & 0 & 0.72(1) & 0.98 & & 0.398(2) &0.196(3) & 0 & 2.4(1) & 0.98(2) &  \\
Fe1 & 4d  & 0 & $\frac{1}{2}$ & $\frac{1}{4}$& 0.40(4) &  0.29 & &
 0 & $\frac{1}{2}$ & $\frac{1}{4}$& 1.56(4) &  0.29(1) &\\
Fe2 & 16i  & 0.1957(6) & 0.0945(5) & 0.2488(7) & 0.40(4) &  0.857 & 3.16(5) &
 0.1972(5) & 0.0938(4) & 0.2507(8) & 1.56(4) &  0.857(6) & 2.46(3)\\
Se1 & 4e & $\frac{1}{2}$ &$\frac{1}{2}$ & 0.136(1) & 0.60(4) &  1  & &
 $\frac{1}{2}$ &$\frac{1}{2}$ & 0.1376(9) & 1.44(4) &  1  &\\
Se2 & 16i & 0.1082(6) & 0.3014(8) & 0.444(3) & 0.60(4) &  1  & &
 0.1098(5) & 0.3006(8) & 0.1451(2) & 1.44(4) &  1 & \\ \hline\hline
\end{tabular}
\end{center}
\end{table*}

Fig.~1(b) shows Bragg peak (101) and (110) of as a function of temperature.
The (110) peak measures the differentiation of the 16$i$ and 4$d$ Fe sites in the $\sqrt{5}\times\sqrt{5}$ order-disorder transition, from equal partial occupation of both sites in the $I4/mmm$ symmetry to the preferred occupation of the 16$i$ Fe2 site in the lower $I4/m$ symmetry. The magnetic (101) peak represents the development of the block antiferromagnetic order \cite{D020830}, and the staggered magnetic moment reaches a large value 3.16(5) and 2.46(3) $\mu_B$/Fe at 50 and 295 K, respectively (Table~\ref{tab2}). Different from the superconducting samples \cite{D020830,D022882}, the $\sqrt{5}\times\sqrt{5}$ Fe order-disorder transition and the block antiferromagnetic transition concur in K$_{0.99}$Fe$_{1.48}$Se$_2$ as the two peaks simultaneously appear at $T_S\approx T_N\approx 500$ K. The $I4/m$ order parameter in Fig.~1(b) also looks different from that of the superconducting samples, caused by another intervening Fe vacancy order of the $Pmna$ symmetry between $T^*$ and $T_{S,N}$, which is signified by the unresolved twin (101) and (011) Bragg peaks in Fig.~1(a).

\begin{figure}
\includegraphics[width=.95\columnwidth]{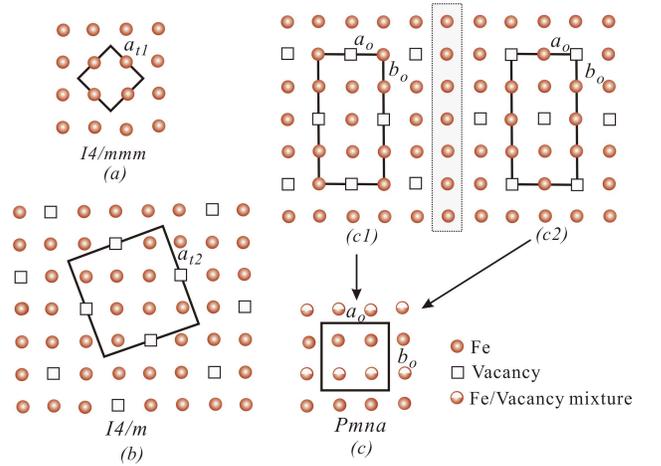}
\vskip -.3cm
\caption{The top view of a Fe layer in the (a) $I4/mmm$, (b) $I4/m$ and (c) $Pmna$ structure, respectively, with the solid line marking the unit cell. With the perfect order shown in (a-c), the sample composition would be KFe$_{2}$Se$_2$, K$_{0.8}$Fe$_{1.6}$Se$_2$ and KFe$_{1.5}$Se$_2$, respectively. The vacancy order in (c) could be an average of those in (c1) and (c2).
}
\label{fig2}
\end{figure}

\begin{table*}[htb!]
\caption{Refined structure parameters of the orthorhombic phase for K$_{0.99}$Fe$_{1.48}$Se$_2$. Space group $Pmna$ with atomic position K1 at $2a$ site (000); 
K2, $2c\, (\frac{1}{2}\frac{1}{2} 0)$; Fe1, $4g\, (\frac{1}{4}\frac{1}{4}\frac{1}{4})$; Fe2, $4g\, (\frac{1}{4}\frac{3}{4}\frac{1}{4}))$; Se1, $4h\, (0\frac{1}{2}z)$ and Se2, $4h\, (00z)$. The occupancy $n$ was fixed at 1 for the Se and K sites, and the $B$ factor constrained to be the same for the same elements in the final refinement to yield $R_p=7.61\%,~wR_p=9.69\% $ and $\chi^2=2.57$ at 350 K; $R_p=5.95\%,~wR_p=7.72\% $ and $\chi^2=3.023$ at 425 K. 
}
\label{tab3}
\begin{center}
\begin{tabular}{c|c|cccc|cc|ccc|cc} \hline\hline
 & phase&  & &  & &
\multicolumn{2}{|c}{$z$} & 
\multicolumn{3}{|c}{$B (\AA^{2}$)} &
\multicolumn{2}{|c}{n } \\ 
T (K) & fraction & a ($\AA$)& b ($\AA$)& c ($\AA$)& V ($\AA^3$) &Se1 & Se2 & K & Fe & Se & Fe1 & Fe2 \\ \hline
350 & 60.0(4)\% & 5.6658(4) & 5.5959(4) & 13.798(2) & 437.47(8) & 0.1514(7) & 0.3582(9) & 2.1(2)  & 1.41(9) &  2.3(1)  & 0.90(2) & 0.49(1)\\
425 & 61.3(4)\% & 5.6787(2) & 5.6084(4)& 13.839(2)& 440.75(4) & 0.1489(9) & 0.3591(9) & 3.1(1)  & 1.88(7)  & 2.50(7)  & 0.93(1) & 0.50(1)\\
 \hline\hline
\end{tabular}
\end{center}
\end{table*}

The refined structure parameters at 350 and 425 K of the orthorhombic $Pmna$ Fe vacancy order are shown in Table~\ref{tab3}. The phase fraction is about 60\% and the 4$g$ Fe2 site is half occupied at the temperatures. The vacancy pattern in the Fe plane is shown in Fig.~2(c) with a $\sqrt{2}\times\sqrt{2}$ supercell. In our refinements, we cannot distinguish between this pattern of a random Fe2 site occupation and the domain average of the two Fe2-ordered patterns with a $2\sqrt{2}\times\sqrt{2}$ supercell in Fig.~2(c1) and (c2). The $2\sqrt{2}\times\sqrt{2}$ vacancy pattern has been previously discussed in work on TlFe$_{1.5}$Se$_2$ and TlFe$_{1.5}$S$_2$ \cite{jmmm86}. A $\sqrt{2}\times\sqrt{2}$ lattice reconstruction has been observed in the TEM study, however, with a $C_4$ symmetry \cite{D012059}, while the $\sqrt{2}\times\sqrt{2}$ structure in Fig.~2(c) possesses a $C_2$ symmetry.

Our sample composition K$_{0.99}$Fe$_{1.48}$Se$_2$ can almost ideally fulfill the Fe vacancy pattern in Fig.~2(c or c1,2). However, the orthorhombic vacancy order is only one of the two competing orders which have been realized in the sample when $T<T_S$. More surprisingly, it is completely replaced below $T^*$ by the competing $I4/m$ vacancy order, which $\sqrt{5}\times\sqrt{5}$ vacancy pattern is ideally compatible with the K$_{0.8}$Fe$_{1.6}$Se$_2$ composition. 
This experimental result can be understood if the combined antiferromagnetic and Fe vacancy order renders the $I4/m$ structure much more energetically favorable than the $Pmna$
structure. Thus, at zero temperature, the lower energy $I4/m$ state stabilizes. At elevated temperature, the mixing in of the higher energy $Pmna$ state can lead to a lower free energy through the entropy term. This theoretic picture is consistent with recent LDA calculations which indicate large energy gain from the block antiferromagnetic order and its associated tetramer lattice distortion \cite{D021344,D022215}.

\begin{figure}
\includegraphics[width=\columnwidth]{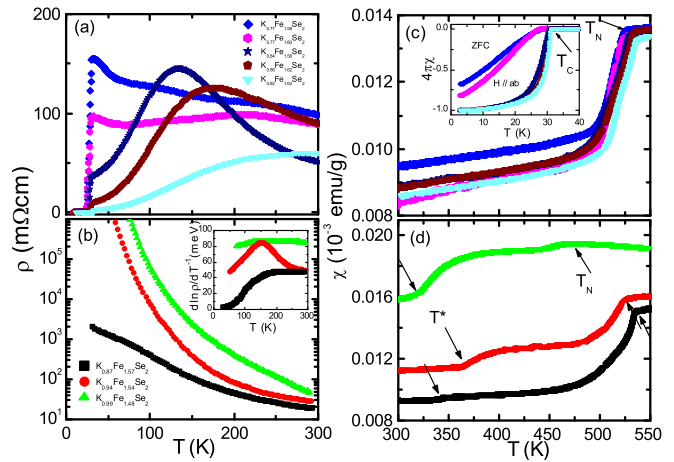}
\vskip -.3cm
\caption{Resistivity of (a) superconducting and (b) insulating samples. The inset to (b) measures the activation gap in transport, which closes when $x$ approaches 0.86. 
Magnetic susceptibility of (c) superconducting and (d) insulating samples. The rapid decrease in (c) corresponds to the rapid development of magnetic order in superconducting samples \protect\cite{D020830}, while the staged decrease in (d) is corroborated by corresponding development of magnetic order in Fig.~1(b).}
\label{fig3}
\end{figure}

The valence of the Fe ion in the insulating K$_{0.99}$Fe$_{1.48}$Se$_2$ sample is also very close to 2+. Actually, all the samples in our K$_{x}$Fe$_{2-y}$Se$_2$ series have the Fe valence close to 2+. Therefore, the sample compositions (black circles on the basal plane in Fig.~4) cluster along the black line $x=2y$ which ensures a 2+ Fe valence for K$_{x}$Fe$_{2-y}$Se$_2$. 
The K$_{0.99}$Fe$_{1.48}$Se$_2$ sample locates on the right side of the phase diagram. The poor degree of the $\sqrt{5}\times\sqrt{5}$ order, constrained by the mismatched number of Fe, is expected to introduce strong impurity scattering of the electrons. The $d \ln \rho/d T^{-1}$ becomes strongly temperature dependent upon moving to the left in the phase-diagram, see inset of Fig.~3(b), and the transport gap closes and $\rho(T)$ approaches logarithmic divergence at K$_{0.87}$Fe$_{1.57}$Se$_2$, see Fig.~3(b). Correlation between magnetic susceptibility in Fig.~3(d) and the microscopic order parameter in Fig.~1(b) can be made. The sharp drop in $\chi(T)$ at $T_N$ and $T^*$ is most likely due to the usual anisotropy gap in spin excitation spectrum.
Henceforth, $T_N$ and $T^*$ can be readily measured using a PPMS (open symbols in Fig.~4). 

\begin{figure}
\includegraphics[width=.8\columnwidth,angle=90]{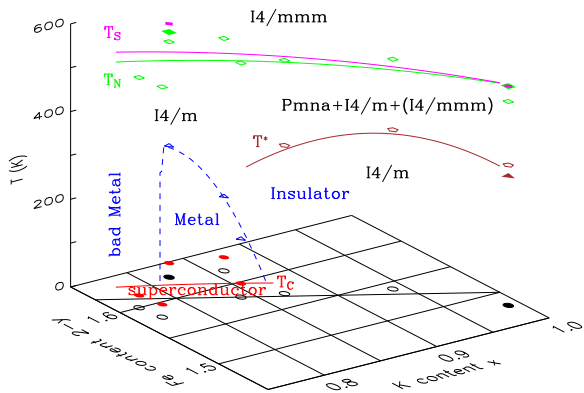}
\vskip -.8cm
\caption{Phase-diagram of K$_x$Fe$_{2-y}$Se$_2$. 
Above $T_S$, the material is in the tetragonal $I4/mmm$ structure with {\em fractional} random occupancy of the Fe site, see Table 1 for details. When the number of Fe $2-y\sim 1.6$, a order-disorder transition to the $\sqrt{5}\times\sqrt{5}$ super-structure [Fig.~2(b), space group $I4/m$] with very high degree of order occurs at $T_S$, see Table 1 in \cite{D020830}. When $2-y$ approaches 1.5, the orthorhombic super-structure [Fig.~2(c), space group $Pmna$, see Table 3 for details] coexists with a partially ordered $I4/m$ structure and the disordered $I4/mmm$ structure below $T_S$; Below $T^*$, only the {\em partially ordered} $\sqrt{5}\times\sqrt{5}$ structure survives, see Table 2.
The $T_N$ remains high for all samples.
Only when perfect vacancy order develops near $2-y= 1.6$, insulator-metal crossover occurs, which is followed by the superconducting transition at $T_c$.}
\label{fig4}
\end{figure}

With further moving towards the ideal sample composition for the $\sqrt{5}\times\sqrt{5}$ Fe vacancy order at $2-y=1.6$, a crossover from insulating to metallic behavior occurs at low temperature, manifesting in the broad peak in resistivity in Fig.~3(a). With the perfecting of the $\sqrt{5}\times\sqrt{5}$ order, disorder scattering is expected to disappear. The two leftmost samples do not complete the crossover to metal and their resistivity stays in the bad metal regime. Superconductivity occurs in this non-insulating phase, see Fig.~3(a,c) and Fig.~4. The three metallic samples
rapidly reaches the Meissner diamagnetic state as demonstrated in the inset to Fig.~3(c) by susceptibility measured with a 10 Oe magnetic field applying along the thin $ab$-plane. The correlation between superconductivity and the normal state transport property in the K$_x$Fe$_{2-y}$Se$_2$ system is very similar to that demonstrated previously for the Fe$_{1+\delta}$Te$_{1-z}$Se$_z$ superconducting system \cite{C035647}, which may be regarded as the $x=y=0$ member of the K$_x$Fe$_{2-y}$Se$_2$ family.

In summary, iron chemistry dictates that $x\approx 2y$ in the K$_{x}$Fe$_{2-y}$Se$_{2}$ system. The iron vacancy orders geometrically demand certain number of Fe ions per unit cell to be perfectly fulfilled. The samples around the K$_{0.8}$Fe$_{1.6}$Se$_2$ composition can realize a highly
ordered $\sqrt{5}\times\sqrt{5}$ structure. The stable energy of the antiferromagnetic and vacancy ordered $\sqrt{5}\times\sqrt{5}$ state preempts the $\sqrt{2}\times\sqrt{2}$ or $2\sqrt{2}\times\sqrt{2}$ iron vacancy order to be a ground state even for samples around the KFe$_{1.5}$Se$_2$ composition.
We identify the degree of the vacancy order, measured by the ratio $n$(Fe2)/$n$(Fe1) in the $I4/m$ structure, as the controlling factor of the metallic and superconducting state in the new iron selenide superconductors.

We thank X. H. Chen, M. H. Fang, N. L. Wang, Z. Y. Lu, T. Xiang and X. Q. Wang for  discussions. The works were supported by the National Basic Research Program of China Grant Nos.\ 2012CB921700,
2011CBA00112, 2011CBA00103 and 2009CB929104 and the National Science Foundation of China Grant Nos.\ 11034012, 11190024, 10974175 and 10934005.


\end{document}